\documentclass[a4paper,11pt]{article}

\usepackage{jinstpub}

\usepackage{float}
\usepackage{xcolor}

\usepackage{siunitx}
\usepackage{tikz}

\usepackage{lineno}

\usepackage{subcaption}
\usepackage{booktabs}
\usepackage{pdfpages}
\title{FAT-GEMs: (Field Assisted) Transparent Gaseous-Electroluminescence Multipliers} 

\author[a]{S. Leardini*}
\author[a]{A. Saá-Hernández}
\author[b]{M.~Kuźniak}
\author[a]{D.~González-Díaz}
\author[c]{C.\,D.\,R. ~Azevedo}
\author[c]{F.~Lucas}
\author[a]{P.~Amedo}
\author[b]{A.\,F.\,V.~Cortez}
\author[a]{D.~Fernández-Posada}
\author[d]{B.~Mehl}
\author[b]{G.~Nieradka}
\author[d]{R.~de~Oliveira}
\author[a]{V.~Peskov}
\author[b]{T.~Sworobowicz}
\author[d]{S.~Williams}

\affiliation[a]{IGFAE, Universidade de Santiago de Compostela, ES} 
\affiliation[b]{Astrocent, Nicolaus Copernicus Astronomical Center of the Polish Academy of Sciences, Rektorska 4, 00-614 Warsaw, Poland}
\affiliation[c]{{Institute of Nanostructures, Nanomodelling and Nanofabrication (i3N), Universidade de Aveiro, Portugal}} 
\affiliation[d]{CERN, Esplanade des Particules 1, Meyrin, Switzerland}

\emailAdd{Sara.Leardini@usc.es}

\abstract{The idea of implementing electroluminescence-based amplification through transparent multi-hole structures (FAT-GEMs) has been entertained for some time. Arguably, for such a technology to be attractive it should perform at least at a level comparable to conventional alternatives based on wires or meshes. We present now a detailed calorimetric study carried out for 5.9~keV X-rays in xenon, for pressures ranging from 2 to 10~bar, resorting to different geometries, production and post-processing techniques. At a reference voltage 5~times above the electroluminescence threshold ($E_{EL,th}\sim0.7$~kV/cm/bar), the number of photoelectrons measured for the best structure was found to be just 18\%~below that obtained for a double-mesh with the same thickness and at the same distance. The energy resolution stayed within 10\% (relative) of the double-mesh value. 

An innovative characteristic of the structure is that vacuum ultraviolet (VUV) transparency of the polymethyl methacrylate (PMMA) substrate was achieved, effectively, through tetraphenylbutadiene (TPB) coating of the electroluminescence channels combined with indium tin oxide (ITO) coating of the electrodes. This resulted in a $\times 2.25$-increased optical yield (compared to the bare structure), that was found to be in good agreement with simulations if assuming a TPB wavelength-shifting-efficiency at the level of WLSE=0.74-1.28, compatible with expected values. This result, combined with the stability demonstrated for the TPB coating under electric field (over 20~h of continuous operation), shows great potential to revolutionize electroluminescence-based instrumentation.}


\begin{document}
\maketitle
\flushbottom

\section{Introduction}

The term electroluminescence (EL) is often used, among others, to refer to the detection principle introduced by Conde and Policarpo in the 60's~\cite{Poli1, Poli2, Poli3}. In its simplest description, within this context, it might be seen as the mechanism by which to enhance the ionization response of a gaseous detector, converting ionization electrons (difficult to count) into photons (easy to count) by means of an electric field. Since the accelerated swarm of electrons experiences different energy thresholds for excitation and ionization, the net result is the appearance of an electric field regime where it is possible to have the former without the latter. In noble gases, specifically, the high VUV-scintillation probability per excited state~\cite{DGD_micro, GiveA2nd} results in strong scintillation yields ($Y$) of up to easily 1000's of~ph/e, yet in the absence of avalanche multiplication~\cite{ELCris, ELBeta}. These are the conditions under which EL exhibits its most remarkable features and finds widespread use: in the avalanche-free regime, as noble gases possess no inelastic degrees of freedom other than excitations, nearly all energy gained by electrons in the electric field~($qV$) turns into scintillation photons, leading to an approximately linear response~\cite{Oliveira}. This can be expressed handily as:
\begin{eqnarray}
& Y & \simeq K (V- V_{th}) \label{eqV}\\
& Y_z/N & \simeq K (E^*- E^*_{th}). \label{eqE}
\end{eqnarray}
The second expression conveys the fact that, when the density-reduced electric field ($E^*=E/N$, a measure of the characteristic electron energy) is above a threshold value needed to excite the medium ($E^*_{th}$), any additional energy gained in the field goes into new excitations. In the above equations $K$ is a proportionality constant, $Y$ refers to the scintillation yield, $Y_z$ to the yield per unit length and $N$ is the number of atoms per unit volume. 
The high efficiency of the EL process has a second consequence, that represents perhaps one of its best known properties: the relative variance of the electroluminescence signal ($Q = (\sigma_Y/Y)^2$) is very small. Its contribution to the calorimetric response of a gaseous detector propagates into the energy resolution (FWHM) as:
\begin{equation}
\mathcal{R} = 2.355 \sqrt{F + Q + \frac{1}{n_{pe}}\left(1+ \frac{\sigma_{_G}^2}{G^2} \right)} \sqrt{\frac{1}{n_e}}.
\end{equation}
In the above expression $F$ represents the Fano factor of the medium (0.1-0.2 in noble gases~\cite{Elena}), $n_{pe}$ is the number of detected photons (`photoelectrons') per electron, and $\sigma_G/G$ refers to the relative width of the single-photon distribution function of the photosensor. The last term involves the number of ionization electrons, $n_e = \varepsilon / W_{I}$, with $\varepsilon$ being the energy deposited in the medium and $W_I$ the average energy needed to release an electron~\cite{Rolandi}. The term $Q$ is generally so small~\cite{Oliveira, Escada} that it has eluded experimental determination except in the presence of electronegative gases such as CO$_2$~\cite{HenriquesCO2}. Due to the smallness of $Q$, as soon as the number of detected photons per primary electron is of the order of $n_{pe}=10$ or more, the energy resolution will approach the intrinsic limit of the gas medium~\cite{Dave}.

The aforementioned characteristics make electroluminescence, nowadays, the workhorse of many experiments based on Time Projection Chambers (TPCs) in the field of Rare Event Searches \cite{Irastorza_RES}, such as XENON~\cite{Xenon-1T}, LZ~\cite{LZ}, PandaX~\cite{PandaX}, and DarkSide~\cite{DSide} (aimed at direct WIMP Dark Matter detection) or NEXT~\cite{NEXT} (aimed at measuring $\beta\beta0\nu$ decay). Its use is foreseen in upcoming experiments such as RED~\cite{RED} and GANESS~\cite{GANESS} (aimed at precision studies of coherent neutrino scattering), and it can be found, too, in applied research for instance in proposals for Compton cameras~\cite{ComptonCamera} and Compton dispersion at next-generation Light Sources~\cite{ComptonDispersion}. In such experiments, the single-electron detection capabilities of electroluminescence are essential to extend WIMP sensitivity down to very low masses (e.g.~\cite{WIMP_LM}), while near-Fano energy resolution is required for precise calorimetry in $\beta\beta0\nu$ searches, thus avoiding contamination from $\beta\beta2\nu$ decays and natural radioactivity~\cite{bb0nu}. High-rate environments like~\cite{ComptonCamera} and~\cite{ComptonDispersion} benefit from the absence of ion feedback and subsequent high rate capability stemming from the avalanche-free nature of the electroluminescence process. 

For all their beauty, the electroluminescence process in presently running experiments is invariably implemented based on early designs relying on wires or meshes, and suffer from the inevitable scaling-up limitations due to electrostatic sagging and deformation~\cite{Leslie} and, chiefly, defects~\cite{Discussions}. During the operation of the NEXT-NEW detector, for instance, the pressure-reduced electric field was 1.2~kV/cm/bar at 10~bar, a mere 0.5~kV/cm/bar above the EL threshold and 3~times below the values achieved in the present work and, generally, on small R\&D setups. Thus, when possible, the commissioning of the EL-region becomes a separate endeavour, requiring purposely-built chambers several square meter in size, essentially as large as the final operating conditions demand~\cite{pancake}. Given that experiments keep increasing in size to achieve sensitivity, it seems very unlikely that the situation will improve except if new production techniques are introduced.

Lately, over the last 5-10~years, there has been interesting progress on the development of optically-enhanced structures, and we follow this lead. Transparent substrates like glass, for instance, have been successfully carved into gaseous electron multipliers~(GEMs), be it by photolitography~\cite{Takeshi1} or sand-blasting~\cite{Kostas1}. They are sturdy and immune to sagging or deformation, and can be easily tiled to cover large areas~\cite{ARIADNE}. At the thicknesses and hole sizes needed for efficient electroluminescence (several~mm), structures based on the highly VUV-reflective Teflon~\cite{Axel} or relying on the transparency of PMMA ~\cite{FATGEM2019} have been manufactured by CNC drilling. In order to achieve VUV-transparency, the solid wavelength-shifter polyethylene naphthalate (PEN) was proposed as a bulk material and successfully machined and operated under Xe and Ar for the first time in~\cite{PENGEM}. A natural continuation of the previous works was to resort to TPB coating of the (PMMA-based) GEM channels, theoretically enabling higher wavelength-shifting yields compared to PEN plates. Aiming at wide range of applications, in particular those related to Rare Event Searches, the core materials of such EL-structures are chosen to be radiopure.

Here we present a comprehensive performance study of different FAT-GEM structures under 5.9~keV X-rays, involving different fabrication procedures, different processing (adding wavelength-shifting coating or an internal reflective layer), different geometries (hole sizes), different gases (xenon and argon), and different pressures (2-10~bar). Comparisons with simulations and with a simple 2-mesh configuration are given too.

\section{FAT-GEMs}\label{sec:FATGEM}
\subsection{Fabrication process}
Results from micropattern gas detectors~(MPGDs) operated in avalanche mode in gaps as small as 50~\textmu{m} (typical GEM thickness) suggest that they can potentially produce a high optical output in a noble gas~\cite{Veloso}. Electroluminescence, on the other hand, benefits from electrified regions where the product of the number density, $N$, times gas gap, $d$, is much larger (up to $P \cdot d \sim 10$ bar${\cdot}$cm at room temperature, compared to $P \cdot d \sim 5\times 10^{-3}$ bar${\cdot}$cm for GEMs at around atmospheric pressure). Geometries with larger values of the product $N \cdot d$ can sustain higher breakdown voltages~\cite{Ben} and therefore provide higher optical gains in avalanche-free conditions (eq. \ref{eqE}). In regard to the electric field configuration, an uniform field region is preferable in terms of intrinsic energy resolution~\cite{Escada} and maximum optical throughput before the onset of multiplication (e.g.~\cite{ELBeta, XenonPaper}). Coincidentally, a high $N \cdot d$ enhances neutral bremsstrahlung radiation below the $E_{EL}$ threshold, too \cite{XeNBrs, ArNBrs}. The main objection against employing large EL gaps is the presence of sagging and deformation when such uniform fields are created by means of meshes or wires suspended on m$^2$ areas and beyond~\cite{Escada, Leslie}.

Thus, in order to create a large-gap, uniform-field, sagging-free structure, PMMA plates of about 5~mm-thickness were procured, drilled at different hole diameters (1-4~mm) on an hexagonal pattern, with a pitch of 5-6~mm. Two additional structures were prepared based on the 2~mm-hole geometry, this time having TPB coated inside the holes. One of them had an additional 3M\texttrademark\ Enhanced Specular Reflector (ESR) layer \cite{esr,weber} placed underneath the cathode. A compilation of the structures studied is provided in table~\ref{tab:structures}.

\begin{figure}[H]
    \centering
   
     \includegraphics[width=13cm]{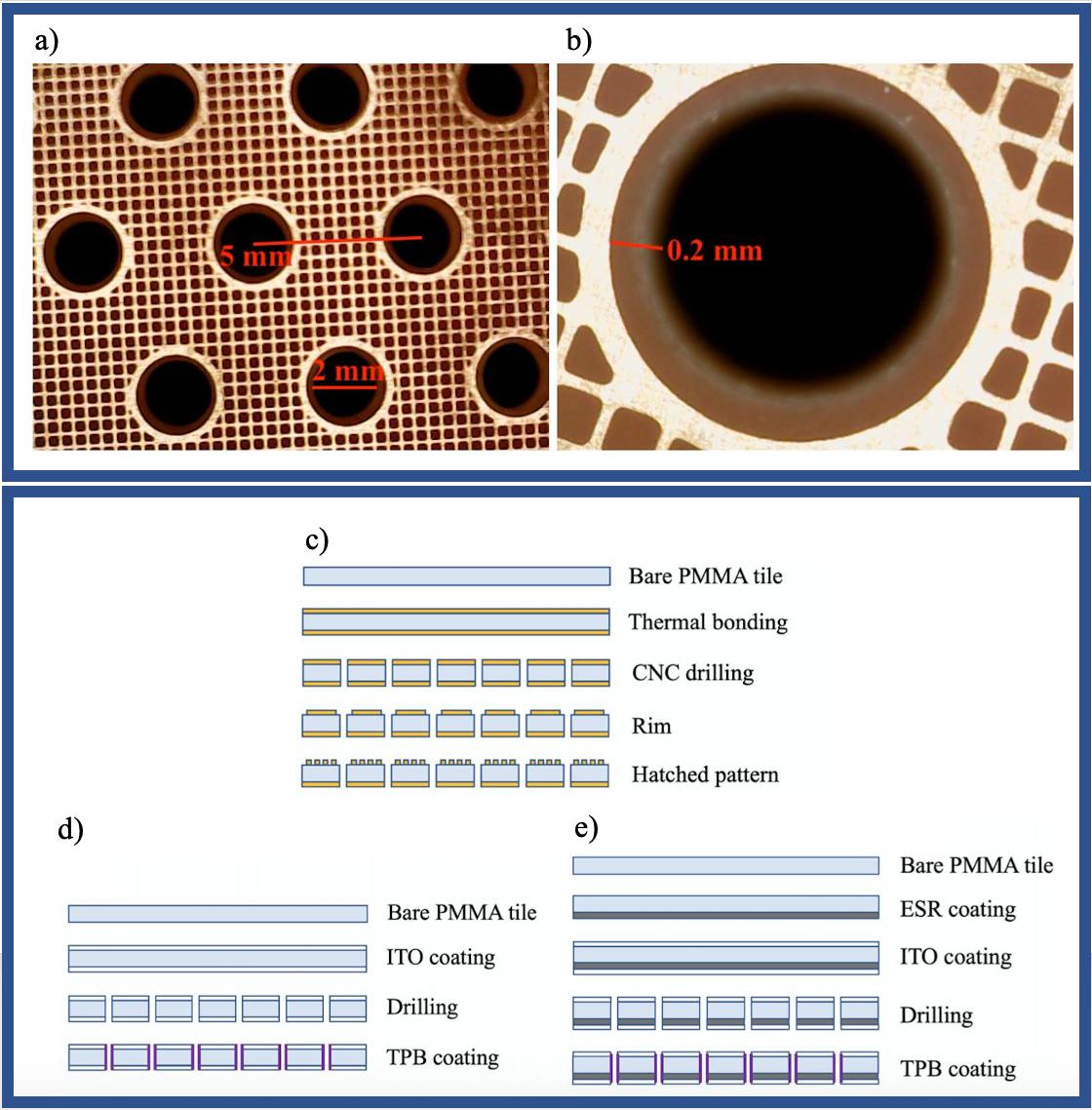}
     
    \centering
    \caption{\footnotesize $a$ and $b$: uncoated FAT-GEM `B' (the hatched-electrode technique was employed in this particular case). $c$, $d$ and $e$: Sketch of the fabrication process for FAT-GEMs produced at CERN ($c$) and at AstroCeNT ($d$ and $e$). FAT-GEMs produced at CERN have thermal-bonded copper electrodes, one of them made semi-transparent through chemical etching of a hatched pattern. For the structures produced at AstroCeNT, ITO electrodes were applied in order to increase the structure transparency; moreover, TPB was coated inside the holes. Procedures $d$ and $e$ differ for the presence of a reflecting ESR layer.}
    \label{fig:fatgemdetails}
\end{figure}

Regular (uncoated) structures were produced in the RD51 workshop at CERN: a bare PMMA plate (7~cm $\times$ 7~cm) was thermally bonded to two circular copper electrodes (6~cm diameter).\footnote{The metallization area was increased with respect to the structures used earlier in~\cite{FATGEM2019}, leaving the active area unchanged, to eliminate possible fringe-fields stemming from charging-up.} Holes were then CNC-drilled, and a rim around them (0.2~mm) was created in a chemical bath, to mitigate corona discharges. Finally, a hatched pattern (inner square side 0.35~mm, trace width 0.1~mm) was made through photolithography on the cathode of the structure. Two things were noticed: on the positive side, electric field simulations showed a negligible change in the electric field compared to the geometry with a solid cathode; on the negative side, thermal-bonding rendered the structure translucent and sub-optimal for further optical treatment, something that was pursued through an alternative fabrication process that is discussed below. Two microscope photos of the resulting structures are provided in Figure ~\ref{fig:fatgemdetails} (a and b), and the fabrication procedures are sketched in Figure ~\ref{fig:fatgemdetails}c.

\begin{table}[ht]
\centering

\begin{tabular}{c| c| c| c| c| c| c}
\toprule
    Descriptor & Hole size (mm) & Pitch (mm) & Thickness (mm) & Number of holes & TPB & ESR  \\ \midrule
    A & 1 & 5 & 4.5 & 30 & No & No \\
    B & 2  & 5 & 4.7 & 30 & No & No \\
    C & 3  & 5 & 4.75 & 30 &No & No\\
    D & 4 & 6  & 4.8 & 18 &No & No \\
    E & 2  & 5 & 5.33 & 31 &Yes & No \\
    F & 2  & 5 & 5.47 &  31& Yes & Yes \\ \bottomrule
\end{tabular}

\caption{Compilation of the FAT-GEM structures characterized in this work.}
\label{tab:structures}
\end{table}

The TPB-coated structures were produced at AstroCeNT (Poland), starting from the application of thin PET films coated with ITO to the two faces of the bare tile, by means of an adhesive film. Compared to the thermally-bonded hatched electrode (transparency 60\%), the PET+ITO resulted in a larger transparency of 79\% (according to the producer~\cite{visiontek}), with the benefit of preserving the optical properties of the plate. Once the PET-ITO films were adhered, holes were made with a manual milling machine equipped with precision drill bits, and finally TPB evaporated inside them (Figure~\ref{fig:fatgemdetails}d). In order to further increase light collection, an ESR layer was interleaved in between the PMMA and the ITO electrode for one of the structures (Figure~\ref{fig:fatgemdetails}e). The thickness of the plates was around 4.7~mm for the uncoated structure and 5.4~mm for the coated ones.

\subsection{Radiopurity}

A mid-size FAT-GEM of 18~cm diameter was purposely-built at the CERN-RD51 workshop and screened mainly to show that the production process did not result in strong contamination. The raw PMMA material was Polycast provided by Spartech (as used in the DEAP-3600 detector), with measured activities as low as 0.11~mBq/kg of $^{235}$U ~\cite{DEAP}. The radiopurity screening was performed at Laboratorio Subterráneo de Canfranc over a practical time of a bit over a month (50.76~days at GeAnayet for the bare PMMA plate, 47.7~days at GeTobazo for the FAT-GEM). The sensitivity of the measurements, presented in Table~\ref{tab:radiopurity}, was mainly limited by the mass of the amplification structure. While competitive radiopurity levels would still need to be demonstrated (especially for a structure that is expected to face the interaction volume), in the light of our current measurements the present fabrication techniques seem promising.

\begin{figure}[H]
    \centering
    \includegraphics[width=13cm]{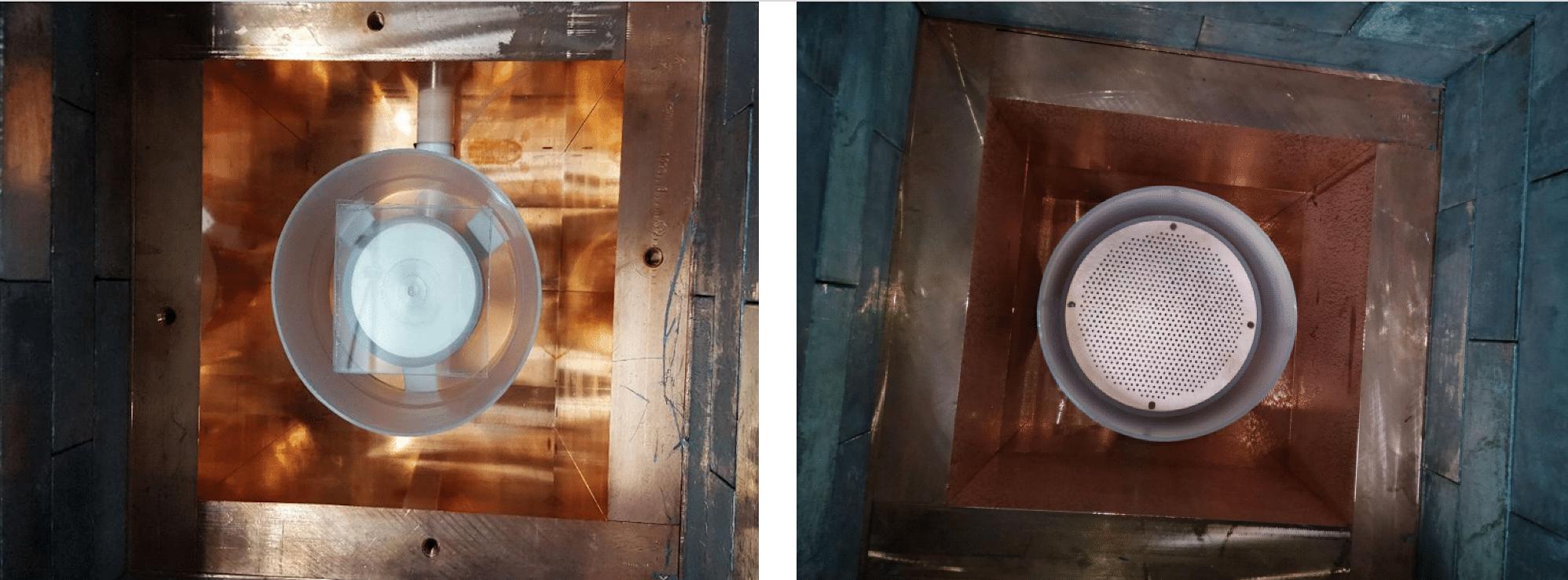}
    \centering
    \caption{\footnotesize Bare PMMA plate (left) and FAT-GEM (right) inside the copper shielding of the setups used for radioactivity screening at Laboratorio Subterráneo de Canfranc.}
    \label{fig:radioactivity}
\end{figure}

\begin{table}[ht]
\centering

\begin{tabular}
{c| c| c| c} \toprule
    Isotope & PMMA (mBq/kg) & FAT-GEM (mBq/kg) [(mBq/cm$^2$)] & PMMA, DEAP-3600 (mBq/kg)\\ \midrule
    U-238/Pa-234m  & $<$340 & $<$791 [$<$0.741] & \\
    U-238/Pb-214  & $<$2.8 & $<$6.9 [$<$0.006]&   \\
    U-238/Bi-214 & $<$2.3  & $<$7.0 [$<$0.007] & \\
    U-238/Ra-226 & & & $<$0.1\\
U-238/Th-234 & & & $<$9.0 \\
    Th-232/Ac-228  & $<$8.8 & $<$22 [$<$0.021] & \\
    Th-232/Pb-212  & $<$2.9 & $<$7.4 [$<$0.007] & \\ 
    Th-232/Tl-208  & $<$6.3 & $<$15 [$<$0.014] & \\
    Th-232 & & &  $<$0.3 \\
    U-235/U-235  & $<$1.9 & $<$6.1 [$<$0.006]& $<$0.6 \\
    K-40 & $<$17  &  $<$38 [$<$0.036] & $<$1.1\\
    Co-60  & $<$0.74 & $<$2.5 [$<$0.002]&  \\
    Cs-137  & $<$1.1 & $<$1.9 [$<$0.002] & \\ 
    \bottomrule

\end{tabular}

\caption{Measurements of radiopurity performed at the Canfranc Underground Laboratory for the bare PMMA (second column) and for a FAT-GEM with thermally-bonded copper electrodes (third column). The last column shows the measurements of the PMMA light guide from the DEAP-3600 detector~\cite{DEAP}. Only upper limits are reported, driven by the sensitivities of the respective assays.}
\label{tab:radiopurity}
\end{table}

\subsection{Working principle}
\begin{figure}[H]
    \centering
    \includegraphics[width=12.5cm]{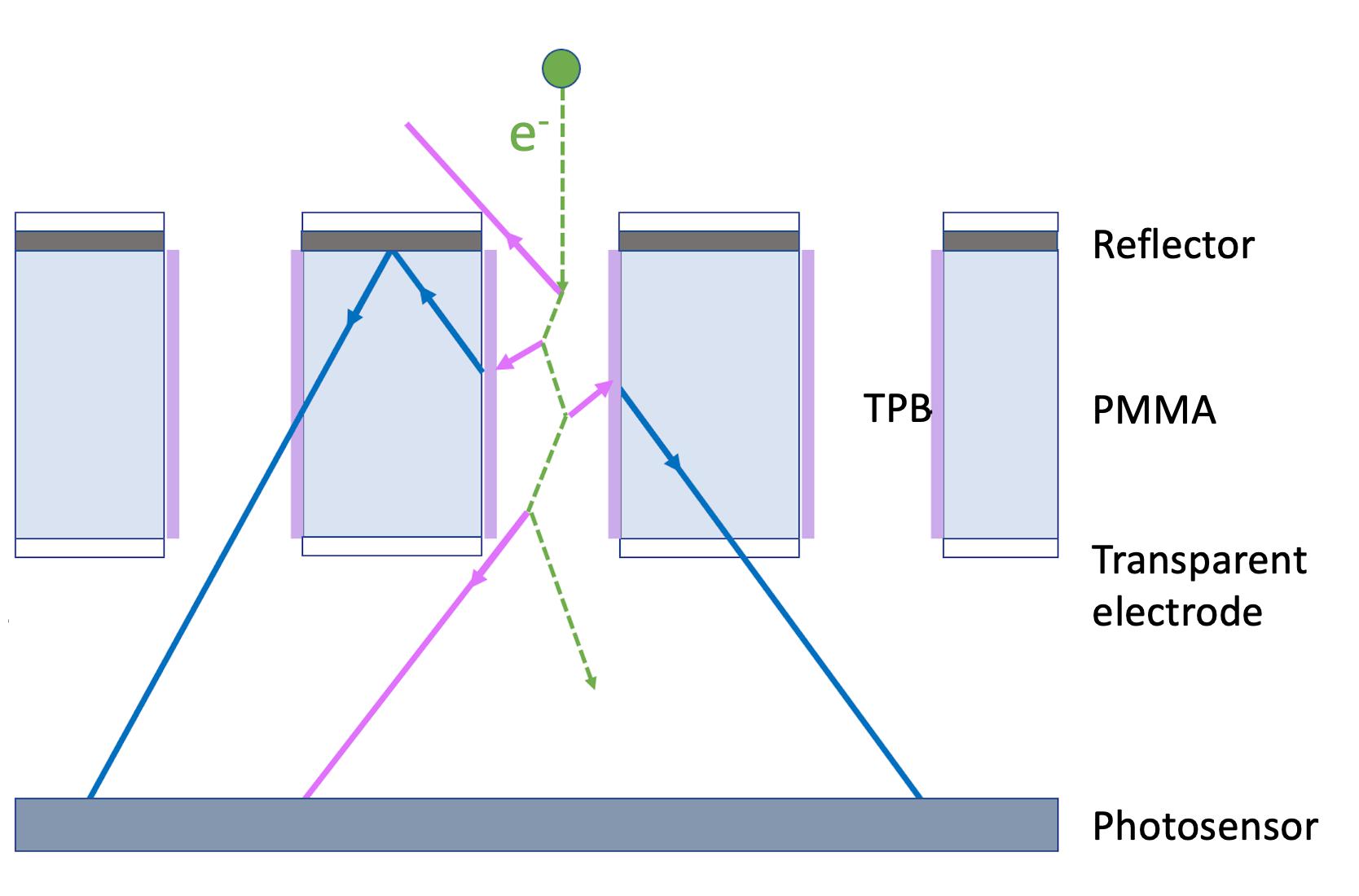}
     
    \centering
    \caption{\footnotesize Illustration of the FAT-GEM concept. The electron enters from above into the channels due to the intense electric field, and produces electroluminescence (VUV). The light is subsequently wavelength-shifted into the visible range through TPB-coating of the holes' walls, reflected at the cathode inner surface and transmitted through the transparent anode until it reaches the photosensor.}
    \label{fig:drawing}
\end{figure}

Having a cutoff typically in the range of 300~nm~\cite{PMMAtransparency}, PMMA is not a VUV-transparent material. Therefore, without further treatment, it is not optimal for use under noble gas scintillation
as the amount of light detected would be limited by the solid angle defined by the holes' walls. In principle, coating the walls with TPB should allow to make use of such a `lost' light at near 100\% efficiency~\cite{tpbeff}, re-emitting it isotropically in a band (peaked at $\lambda\sim$420~nm, with a cutoff around $\lambda\sim$350~nm~\cite{TPBspectrum}) to which the PMMA is fully transparent. Additionally, a reflective layer would help recovering part of the wavelength-shifted light that is emitted in the opposite direction with respect to the light sensor (Figure~\ref{fig:drawing}).

\section{Experimental methods}\label{sec:setup}

\subsection{Setup}
\label{sub:purity}
\begin{figure}[H]
    \centering
    \includegraphics[width=14cm]{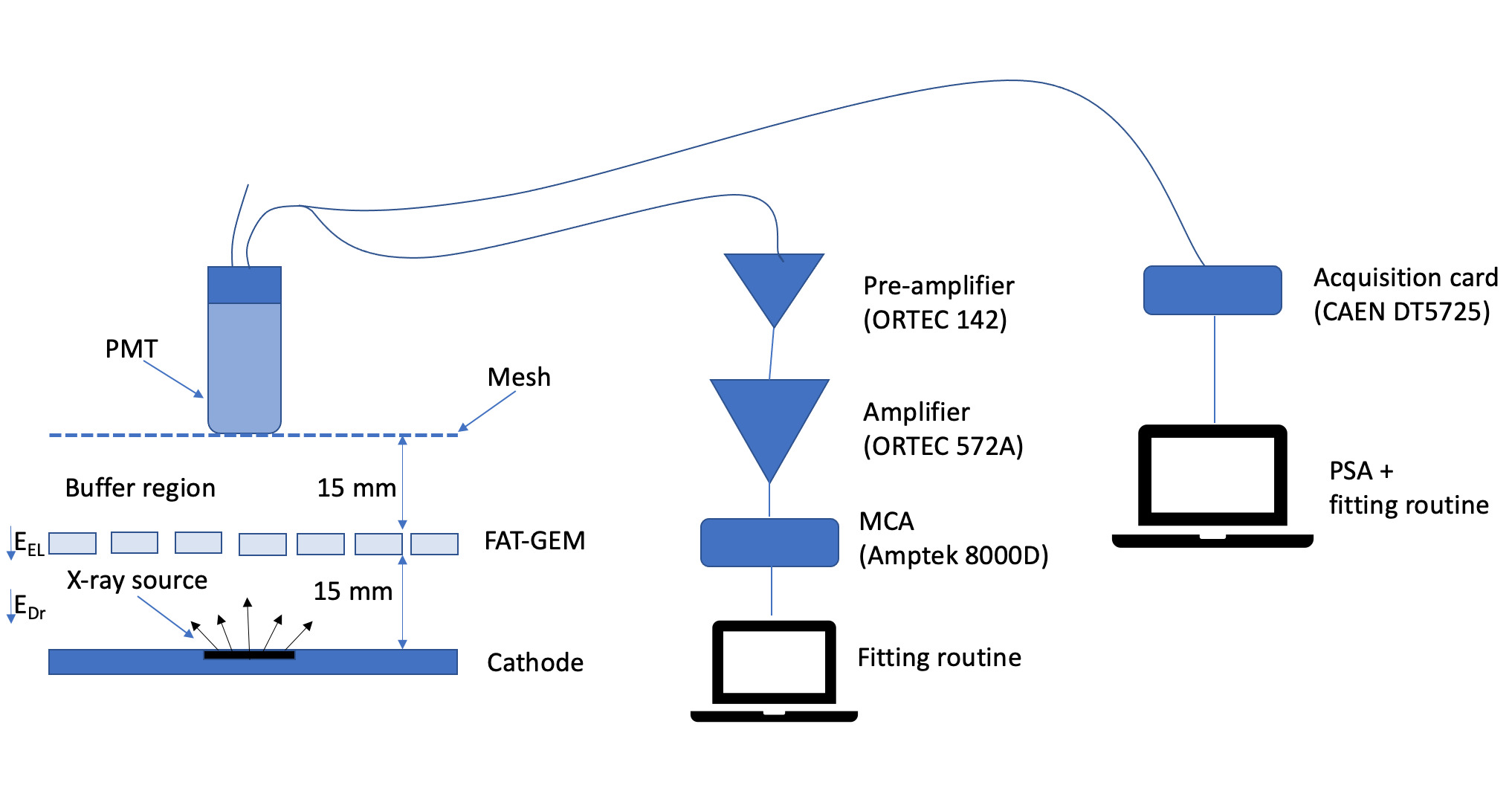}
     
    \centering
    \caption{\footnotesize Sketch of the experimental setup and the two acquisition modes employed during the measurements. The central one was used regularly for most measurements, while the right one was employed to calibrate from MCA channels to photoelectrons.}
    \label{fig:setup}
\end{figure}

The setup used was the same as in previous works~\cite{FATGEM2019, PENGEM}: the electroluminescent structures served as the anode of a drift/conversion region of 15~mm in length, closed by an aluminum cathode with an x-ray source ($^{55}$Fe) placed behind it (Figure ~\ref{fig:setup}). Initially, a high-transparency mesh was placed covering the anode of the amplification structures, but it was removed once assessed, comparing the yield curve of the same structure with and without it, that it had a negligible impact on the performance. 
At 15~mm from the FAT-GEM anode, a Hamamatsu R7378 PMT was placed, covered with a grounded mesh to eliminate fringe-fields from the buffer region leaking into the PMT vacuum. The PMT was then connected to a pre-amplifier (ORTEC~142), an amplifier (ORTEC~572A) and finally to an MCA (Amptek~8000D) to collect the spectra. In order to convert from MCA channels to photoelectrons, a calibration run was performed by connecting the PMT directly to an acquisition card (CAEN~DT5725), as described later.

A sketch of the experimental setup and a description of the acquisition system can be seen in Figure ~\ref{fig:setup}. The container vessel was connected to a pressure/vacuum system and, before filling, it was pumped down to 10$^{-4}$~mbar. During operation the gas was circulated through a cold getter (Entegris~GPU80) specified to trap H$_2$O and O$_2$, with a KNF compressor (N286.15) at around 20~Nl/min.\footnote{Normal liters per minute.}

During the argon measurements, high-purity fresh gas (6N bottle, i.e., ppm-level contamination) was used for each run (namely, for each EL-field series). During the measurements with xenon, on the other hand, the gas was cryorecovered after each run and residual impurities were pumped away. Variations in the yields were observed during the measurements and estimated to typically represent a 15\%. They were attributed to gas contamination and PMT drifts.

\subsection{Data taking and analysis}
\begin{figure}[H]
    \centering
    \includegraphics[width=14cm]{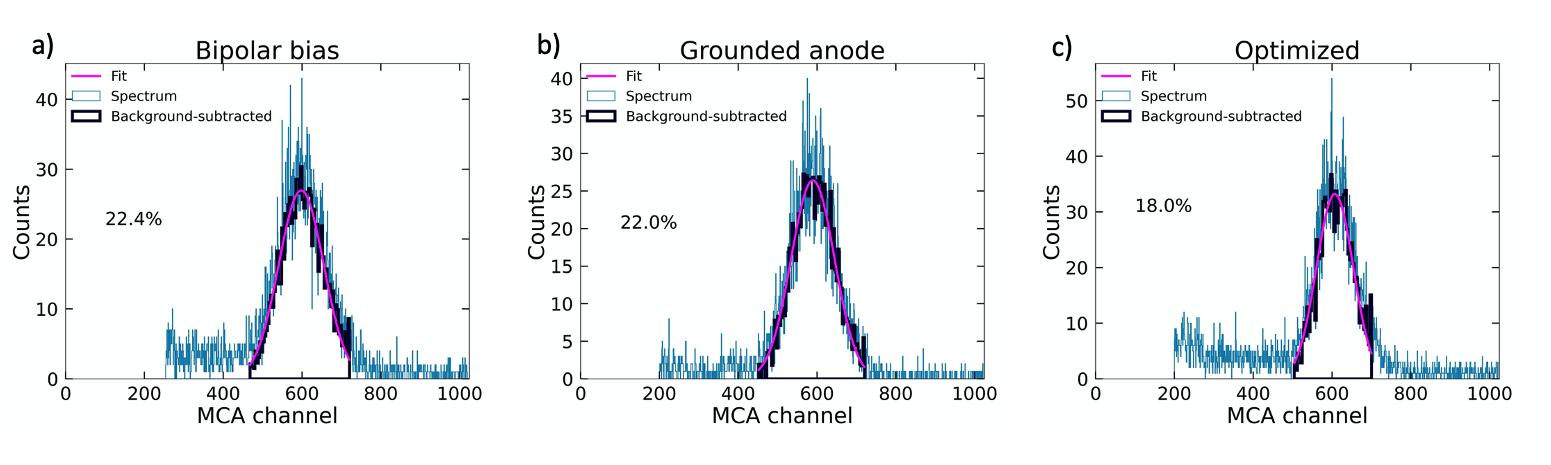}
     
    \centering
    \caption{\footnotesize a) and b): exemplary spectra taken with the MCA (2~mm structure, $P$=2~bar, $E_{Dr}$=167~V/cm/bar, $E_{EL}$ = 4.6~kV/cm/bar), with a Gaussian fit superimposed (22.4\% FWHM -a), and 22.0\% FWHM -b), respectively). This run corresponds to a control measurement performed with bipolar bias (a) and grounded anode (b). The difference in background activity is attributed to the buffer region and was subtracted during analysis. c): MCA spectrum taken for the 2~mm structure at $P$=2~bar, $E_{Dr}$=167~V/cm/bar, $E_{EL}$ = 6.9~kV/cm/bar, corresponding to an energy resolution of 18\% FWHM. The PMT voltage was lower for this point than for the spectra in a) and b), in order to match the MCA range.}
    \label{fig:spectra}
\end{figure}

For each structure, the electroluminescence field ($E_{EL}$) was kept fixed while scanning the drift field ($E_{Dr}$), to achieve the highest electron transmission (a detailed explanation is provided in Sec. \ref{sec:res1}). With the drift field fixed at that value, a scan for different electroluminescence field values was performed. MCA spectra were then stored, rebinned, background-subtracted and finally fitted to a gaussian (Figure~\ref{fig:spectra}). Apart from the electronic noise at low channel count, there was a second source of background on the left hand-side of the spectra correlated with activity in the buffer region (between the structure and the PMT-mesh) when the anode of the FAT-GEM was positively biased (Figure~\ref{fig:spectra}-a). 
Although the size of the buffer region was chosen such that the electric field there was above the electroluminescence threshold at all times (maximum value of 533~V/cm/bar), the background was seen to increase with pressure, pointing to the presence of a scintillation mechanism different from electroluminescence. The only available phenomenon outside corona effect stemming from the connection points (not expected as those were smoothed and screened with black tape), is the direct radiative emission of the drifting electrons in the form of neutral/dipolar bremsstrahlung \cite{XeNBrs}. 
This source of background was assessed by a dedicated run with the 2~mm-hole structure, taking spectra at the same drift and electroluminescence fields, but with the anode of the structure grounded (Figure~\ref{fig:spectra}-b). Figure ~\ref{fig:spectra}-c shows for reference a spectra taken close to the optimal electric field (maximum transmission and highest electroluminescence field). At the highest pressures, when bipolar biasing of the FAT-GEMs was unavoidable due to power supply limitations, the background subtraction was implemented to extract the energy resolution.\footnote{The analysis procedure is as follows: i) an iterative gaussian fit was performed, over a running window defined from the width of the $^{55}$Fe energy distribution; ii) once the algorithm converged to the estimates of the peak position and width, the background was defined as any entry outside a $\pm 2.5\sigma$ band around the peak position; iii) the background was then fitted to a straight line and the interpolated value subtracted from the energy distribution; iv) the gaussian fit was repeated, providing the final values. Studies performed in equivalent conditions, with and without background, allowed to assign the uncertainty of the procedure to be about 1\% (relative), that is included in the uncertainty of the energy-resolution data.}

\begin{figure}[H]
    \centering
    \includegraphics[width=13cm]{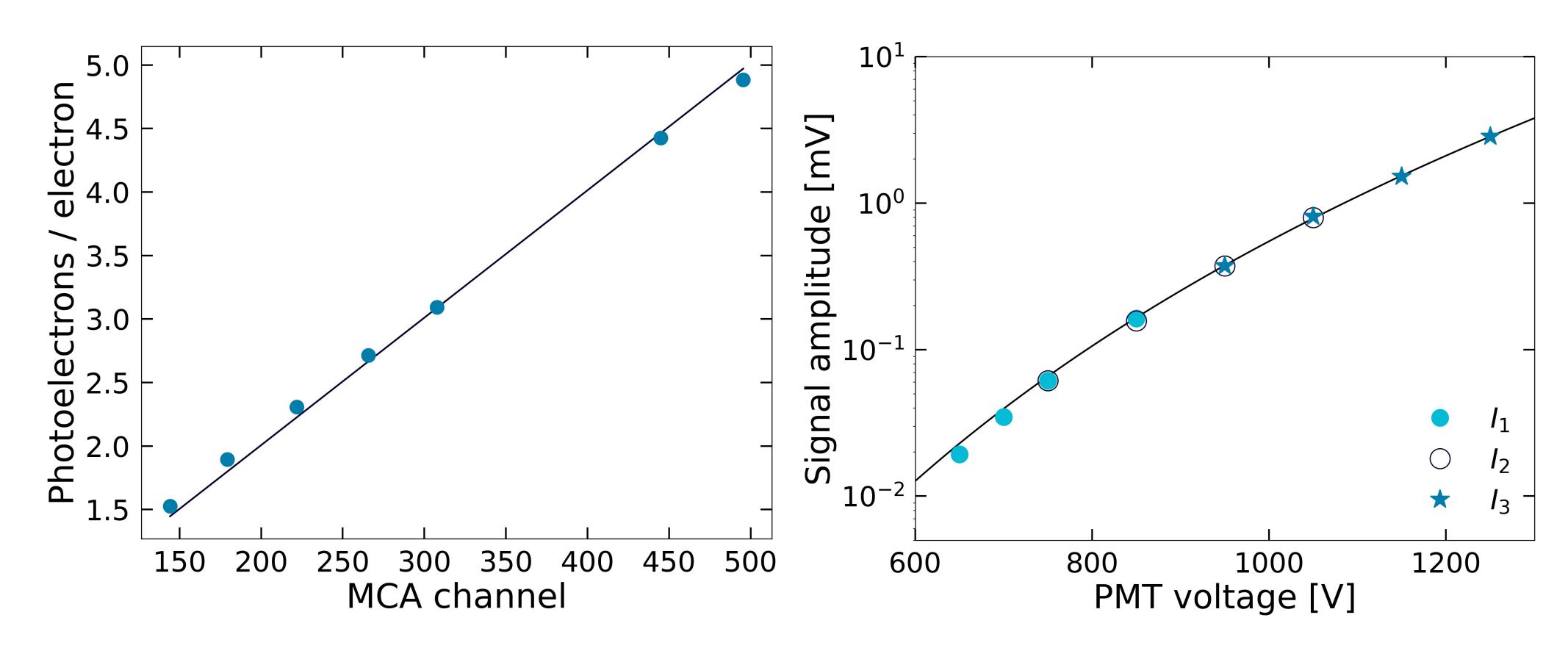}
    
    \centering
    \caption{\footnotesize Left: calibration of the MCA spectra for the 2-mm uncoated structure: $P=8$~bar, V$_{PMT}$=550~V. The $y$-axis shows the $^{55}$Fe X-ray peak position obtained from pulse shape analysis, and the $x$-axis the one from the peak obtained from the MCA. The best-fit parameter is 0.01 photoelectron/electron per MCA channel at V$_{PMT}$=550~V. Right: average amplitude of the PMT signals as a function of the bias voltage for three LED intensities. Each LED intensity series was taken so as to keep the PMT charge per pulse within the same range, and re-scaled afterwards.}
    \label{fig:calibrations}
\end{figure}

In order to convert from MCA channels to photoelectrons, a two-step procedure was followed: first, the PMT was calibrated according to the procedure described in~\cite{XenonPaper}: a green LED was powered with a fast pulser (Agilent~81130A), adjusting the bias voltage until it resulted in a light intensity around the single-photon level. Data was then taken with an acquisition card (CAEN~DT5725) and fitted with gaussians, whose integral values were bound by Poisson statistics. Second, a dedicated run with the 2-mm uncoated structure was performed, setting the same values of drift and electroluminescence fields as for the data taken with the MCA. Through a pulse-shape analysis and fitting routine, and resorting to the single-photon calibration, the number of photoelectrons/electron was obtained. Finally, the slope between the number of photoelectrons/electron and the peak position in the MCA data was found by performing a linear fit, providing the calibration value (Figure~\ref{fig:calibrations}, left).

Since different PMT biases were used during the measurements, the relation between gain and voltage was assessed with an LED. The data was then fitted to a power-law, showing agreement with the data-sheet from the manufacturer (Figure~\ref{fig:calibrations}, right).

\section{Simulations}\label{sec:sim}
\begin{figure}[h!!!]
    \centering
    \includegraphics[width=12.5cm]{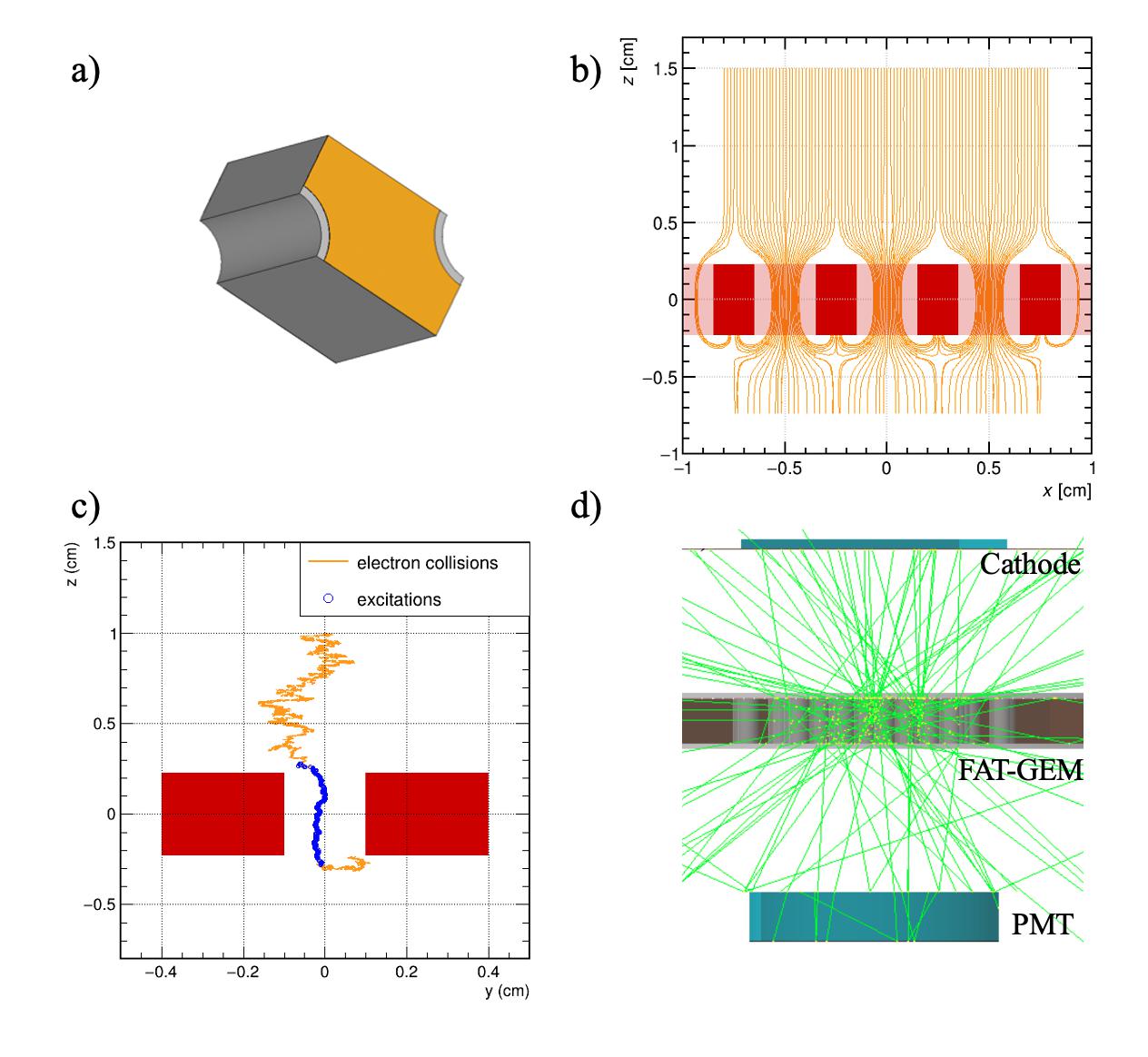}
    \centering
    \caption{\footnotesize Simulations steps: (a) elementary cell in Ansys, (b) Ansys field maps loaded in Garfield++, (c) electron tracking in Garfield++, and (d) photon-tracing in Geant4.}
    \label{fig:simulations}
\end{figure}

Simulations were performed according to the following steps. First, the electric field maps ensuing from an elementary amplification cell were created with Ansys~\cite{ANSYS}, (Figure~\ref{fig:simulations}a).
Field and geometry maps were then imported into Garfield++~\cite{GARFIELD} and extended in the $x$ and $y$ axis (axes parallel to the electrode plane), building the FAT-GEM structure (Figure~\ref{fig:simulations}b).
The position of the primary electrons was generated following an uniform $x$, $y$ distribution centered at the middle of the detector, extending over a $\pm 5$~mm region and placed at 0.1~mm from the cathode plane. The impact of considering a distributed ionization source corresponding to the X-ray mean free path (around 2.7~mm/bar) was evaluated but the differences dimmed too small.
In Figure~\ref{fig:simulations}c the path of an electron is shown: as expected, most of the collisions that produce an excited state happen inside the FAT-GEM hole.

The Garfield++ simulation provided the 3D-distribution of excited states over the entire array of holes. This was then imported into Geant4 \cite{GEANT4} and used to sample the initial positions of the scintillation photons, that were subsequently launched isotropically.\footnote{A control simulation considering excited states produced only in the central hole yielded little differences.} Photon ray-tracing was handled by a Geant4 model with optical properties implemented for all materials, extending the model used in Ref.~\cite{PENGEM}.  New wavelength-dependent optical properties added to the model included the transparency of ITO (approx.~79\%, based on the suppliers' specification~\cite{visiontek}), measured reflectivity for the laminated ESR foil, and TPB properties as in~\cite{2pac}. This allowed us to estimate the geometrical efficiency of the experimental setup (Figure~\ref{fig:simulations}d). A two-mesh configuration was simulated using similar techniques, except that the mesh transparency was introduced as an effective number and the reflectivity of the aluminum cathode \cite{alu} was included too (yielding about a 10\% increase, an effect negligible for FAT-GEMs).

Finally, photon detection efficiency (PDE) curves were considered for four illustrative sensors: Hamamatsu R7378~\cite{pmtqe} (employed in these measurements), R11410-21 (used in XENON1T~\cite{xenonpm}) PMTs, and two representative SiPMs: FBK NUV-HD-SF~\cite{fbk} and Hamamatsu VUV4 (S13370)~\cite{vuv4}.

\section{Results}\label{sec:res}

\subsection{Untreated FAT-GEMs}\label{sec:res1}

\begin{figure}[H]
    \centering
    \includegraphics[width=14cm]{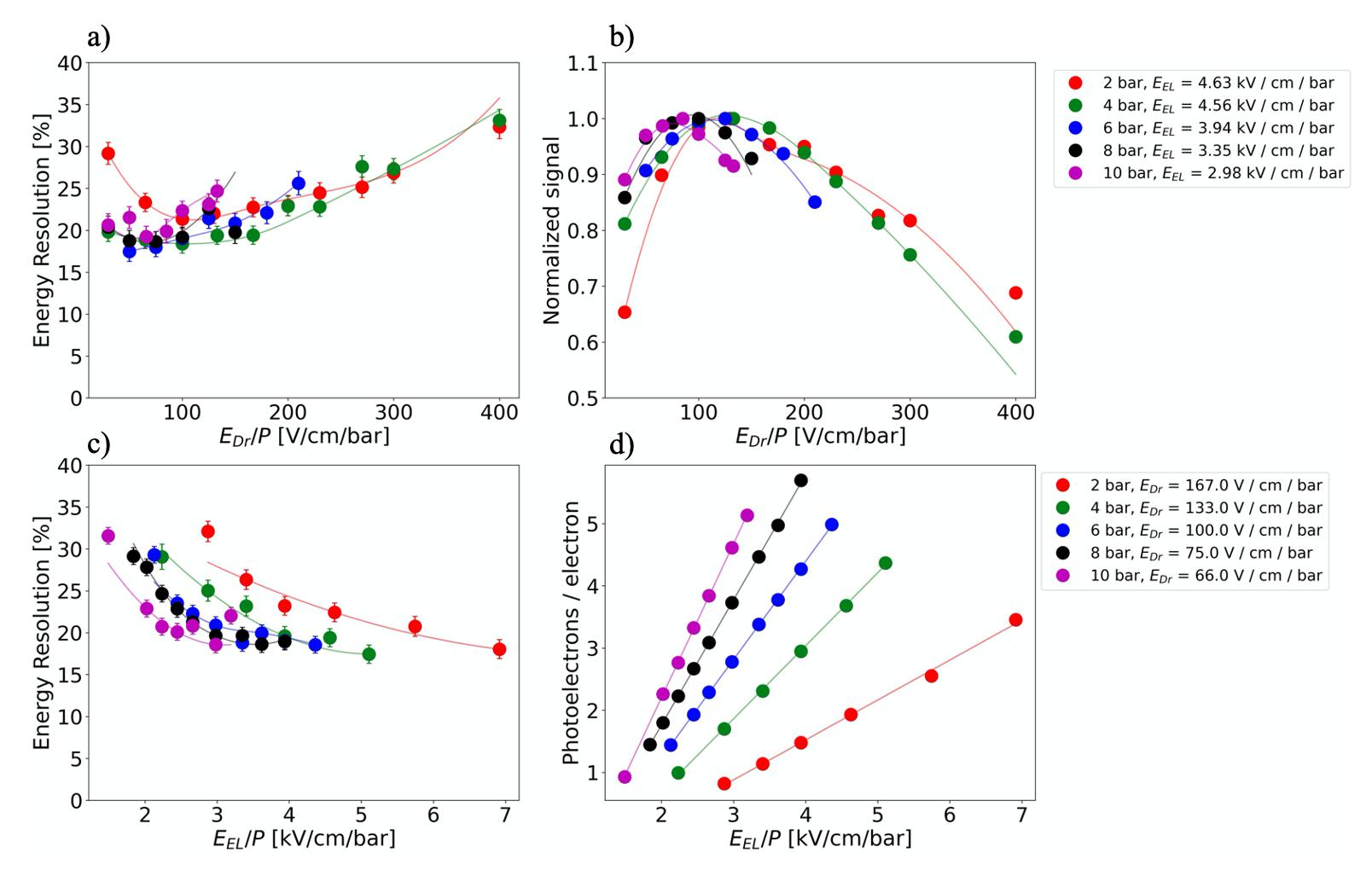}
     
    \centering
    \caption{\footnotesize Drift field ($a$ and $b$) and EL field ($c$ and $d$) scans for the uncoated 2~mm-hole FAT-GEM. The lines for the energy resolution and the transmission curves are splines meant to guide the eye.}
    \label{fig:2mmresult}
\end{figure}

\begin{figure}[H]
    \centering
    \includegraphics[width=8cm]{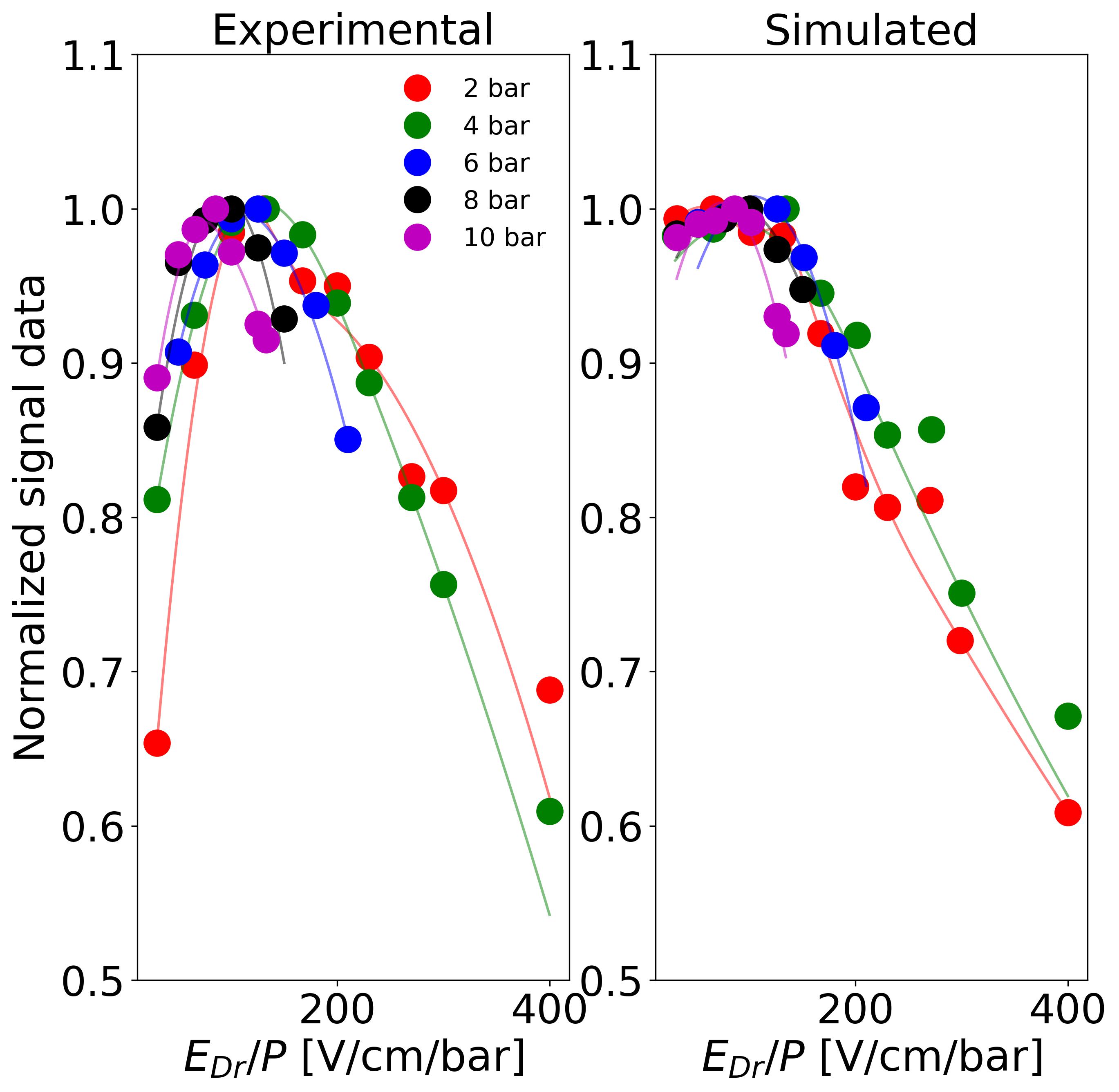}
     
    \centering
    \caption{\footnotesize Experimental (left) and simulated (right) electron-transmission curve for the uncoated 2-mm-hole FAT-GEM. The decrease of transmission is reproduced at high fields and partially at low fields. An extra contribution is suspected to come from attachment to impurities.}
    \label{fig:2mmtransmission}
\end{figure}

In Figure~\ref{fig:2mmresult}, the characterization of the uncoated 2~mm-structure is reported. Data were taken with xenon gas at 2, 4, 6, 8, and 10~bar. The plots labeled as a and b refer to the drift-scan measurements: on the left the energy resolution (FWHM), on the right the light yield normalized to the maximum (`transmission' curve). The decrease of transmission at high drift fields can be explained by the loss of field-focusing: the electron, under a strong drift field, hits the surface of the structure instead of being channeled into the holes. The decrease of transmission at low fields is qualitatively reproduced by simulations too (Figure \ref{fig:2mmtransmission}) and can be attributed partly to electroluminescence extending outside of the hole region and increasing as the drift field increases. However the much stronger trend in data suggests the presence of additional contributions (presumably due to attachment to impurities). 

Figure~\ref{fig:2mmresult}(c and d) shows the results of the electroluminescence scan: on the left the energy resolution, on the right the detected number of photons/electron, namely photoelectrons/electron ($n_{pe}$). One can see that the scintillation yield increases with pressure as expected, except for 10~bar where the voltage limit of our power supplies is reached. The voltage drop across the FAT-GEM amounts to 15~kV in those conditions.

\begin{figure}[H]
    \centering
    \includegraphics[width=12.5cm]{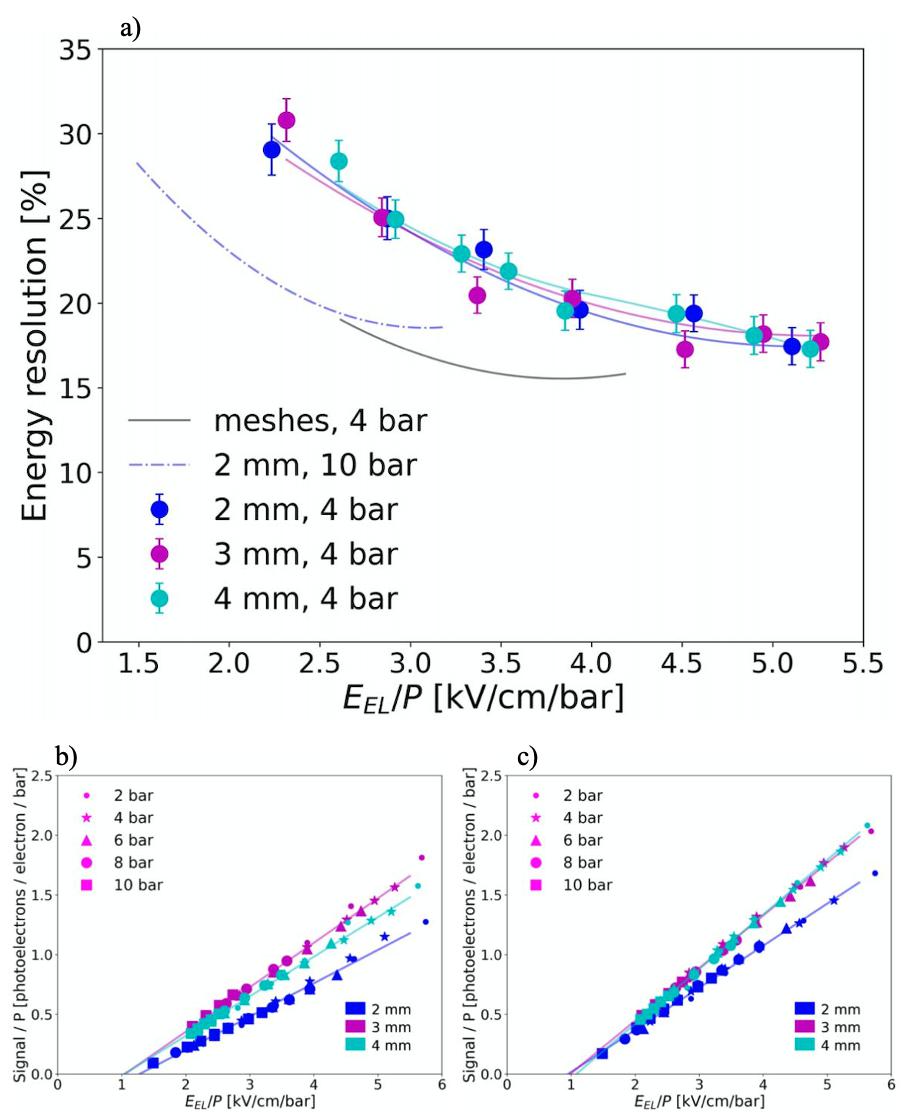}
    \centering
    \caption{\footnotesize $a$: energy resolution at 4~bar for FAT-GEM structures of different hole dimensions (blue - 2~mm `B', purple - 3~mm `C', light blue - 4~mm `D'). For reference, the splines corresponding to the double-mesh data (black, continuous) and 10~bar data (blue, dot-dashed) are given too (and data-points can be found elsewhere in this document). $b$: pressure-reduced scintillation yields measured for FAT-GEM structures of different hole dimensions (dark blue 2~mm - `B', purple 3~mm - `C', light blue 4~mm - `D'). $c$: pressure-reduced scintillation yields simulated for FAT-GEM structures of different hole dimensions. (Measurements and simulations performed for drifts fields corresponding to the transmission plateau).}
    \label{fig:energyres}
\end{figure}

The `typical' best energy resolution was 17-17.5\%, obtained for virtually any structure and pressure, at high E$_{EL}$ fields (Figure~\ref{fig:energyres}, a). A dedicated run with xenon at 4~bar and meshes (electroluminescence gap of 5~mm) yielded an energy resolution a bit over 15\% despite the much increased scintillation (the mesh results are shown through the continuous line in Figure ~\ref{fig:energyres}, a).

In Figure \ref{fig:energyres} (b) the pressure-reduced yields are shown for the three uncoated structures and all pressures. All points fall on a line, showing that operation takes place in pure EL-mode, i.e., largely free from avalanche multiplication. Comparing the results with simulations (Figure \ref{fig:energyres}, c), the latter predict slightly larger yields: 15\% (2~mm), 22\% (3~mm), 30\%~(4~mm). Remarkably, both in simulation and measurements, the threshold to have scintillation light is observed to be 1~kV/cm/bar (above the canonical 0.7~kV/cm/bar for uniform-field conditions in xenon \cite{XeNBrs}) for all structures. The increase can be interpreted as the lower effective field in the central axis of the hole (where electrons are field-focused) compared to a parallel-electrode configuration. In data, a small deviation is observed for 2~mm structures, providing a hint that the yields at high $E_{EL}/P$ and low $P$ might have a small contribution from avalanche multiplication, biasing the linear fit towards higher threshold values.

An additional structure with 1~mm hole size 
was characterized, yielding a factor $\times 3$ less light than the 2~mm structure. This seems to support the expectation from simulation, where a minimum diameter size of 1.75~mm is needed in order to reach 100\% `entrance $\times$ transmission' probability for the impinging electron, on a 5~mm-long xenon channel at typical EL fields.

\subsection{TPB-coated FAT-GEMs}
\begin{figure}[H]
    \centering
    \includegraphics[width=12.5cm]{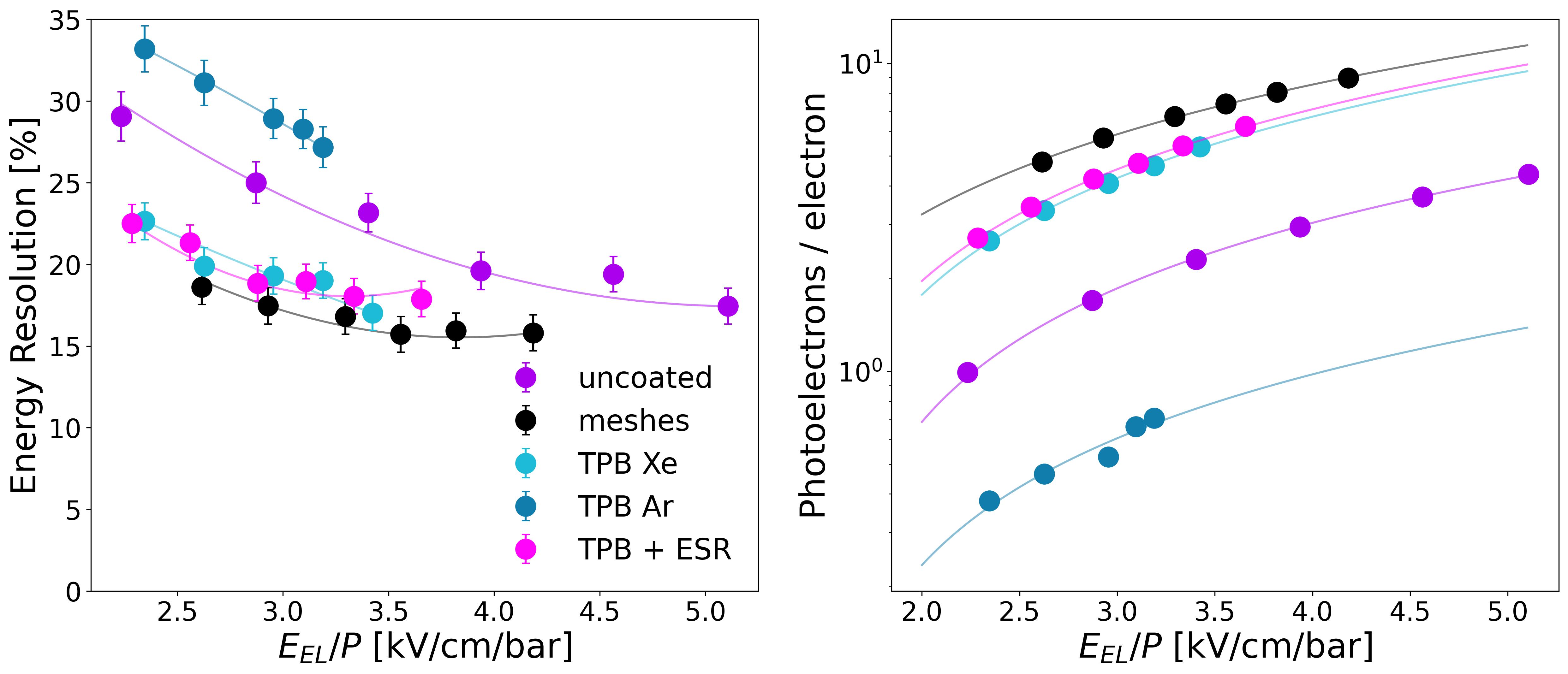}
     
    \centering
    \caption{\footnotesize Performance studies of the TPB-coated FAT-GEMs (light and dark blue - `E', pink - `F'), including results for the uncoated structure (`B'), and two parallel meshes (black circles). These studies were performed at 4~bar, in xenon and argon.}
    \label{fig:tpbresults}
\end{figure}

A direct assessment of the impact of TPB at the holes' walls was made for the 2~mm structure under argon, since our PMT has a fused-silica window and is therefore blind to its scintillation. As expected, no signal was observed before coating with TPB, however a clear spectrum was formed after doing so (Figure~\ref{fig:tpbresults}, dark-blue circles).

A run with xenon at 4~bar showed a 125\% average increase of detected light with respect to the uncoated structure, resulting on a mere 25\% deficit with respect to meshes at the maximum operating field of 3.5~kV/cm/bar (Figure~\ref{fig:tpbresults}, light-blue circles). Such a field was significantly lower than the 5~kV/cm/bar applied to the uncoated structures and was chosen simply to fully preserve the structure for further studies, avoiding risk of discharges and TPB damage. At that field, the energy resolution was comparable with the one provided by the two-mesh setup (Figure~\ref{fig:tpbresults}, left).

A structure with ESR was also tested in xenon at 4~bar (pink circles), providing just 6\% more light than the coated structure without reflector. As discussed later in text, such a modest increase can be understood largely due to the presence of internal reflection inside the PMMA, due to the specular-reflection characteristics of ESR, and/or a bad optical coupling. A path to improve this performance is discussed later in text.

\begin{figure}[H]
    \centering
    \includegraphics[width=12.5cm]{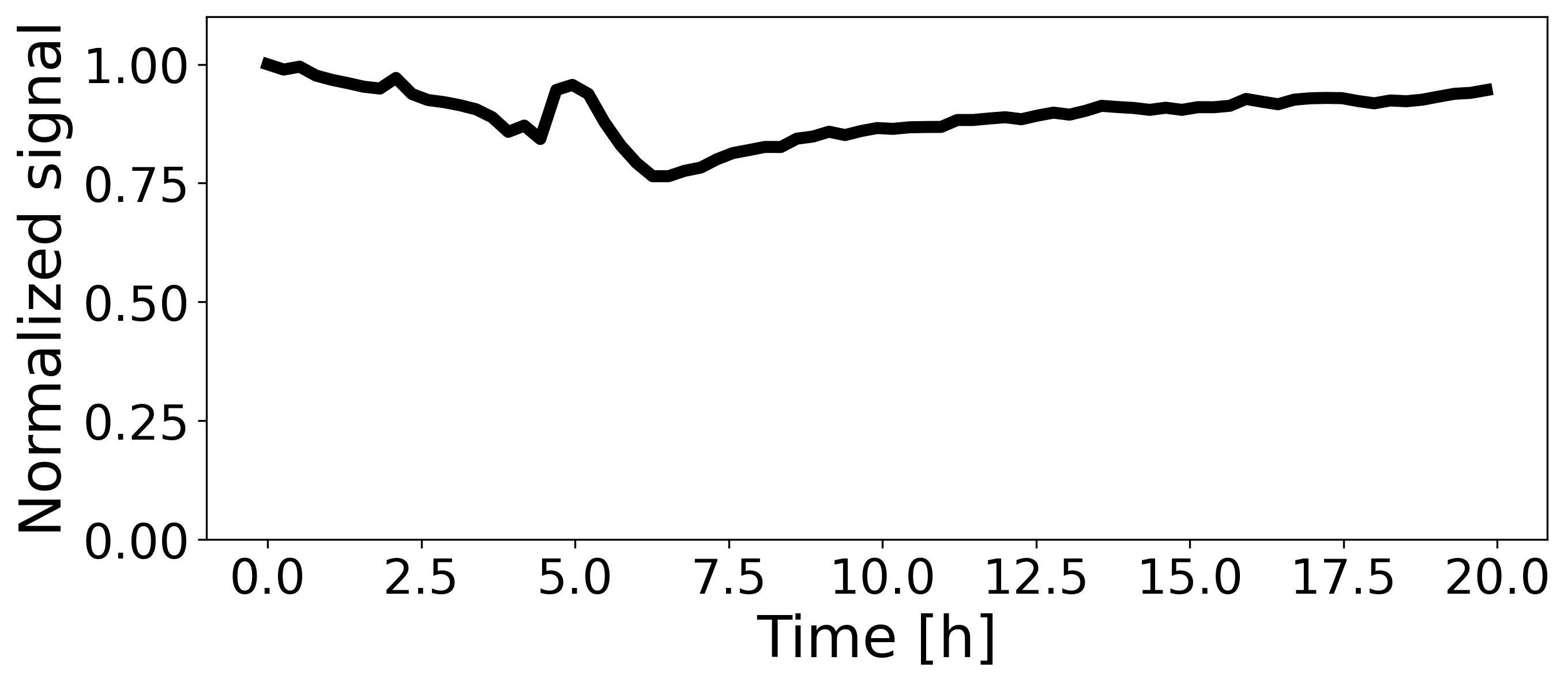}
     
    \centering
    \caption{\footnotesize Study of possible TPB-damage induced by sustained operation under high voltage ($\Delta{V}=6.8$~kV). The optical gain of the FAT-GEM was monitored over 20~h in xenon at 4~bar and, except for a non-monotonous variation (attributed to the PMT drift), no strong short-term deterioration could be seen.
    }
    \label{fig:tpbaging}
\end{figure}

Short-term aging of the TPB-coated FAT-GEMs was studied in xenon at 4~bar, operating the structure with $\Delta V$=6.8~kV continuously over 20~hours (event rate $\sim$5~Hz). Non-monotonous variations within a maximum of 25\% were observed, attributed mainly to PMT drifts. The strong correlation with the temperature variation during a day-night cycle suggests that this transient behaviour arises from small variations in the electrical properties of the passive components of the base (Figure~\ref{fig:tpbaging}). No indication of permanent damage was evident after the test.

\section{Discussion} \label{sec:dis}

The best energy resolution that was measured for FAT-GEMs in this work was 17\% (FWHM), at the 5.9~keV peak of $^{55}$Fe.  This represents almost a factor two increase with respect to the value expected from the Fano factor alone~\cite{FANO1, FANO2, FANO3, FANO4, FANO5, FANOLAST}.
The measurement is not photon statistics -limited, as similar asymptotic values were obtained for different optical gains (up to a number of detected photoelectrons per electron $n_{pe} = 10$). Further, similar results were obtained both with a double-mesh and with a drift-less configuration (the latter being more immune to attachment). The deviation from the Fano limit can be qualitatively understood from the afterpulsing of this PMT model (as observed during the LED characterization): it appears with a delay of about 250~ns, and can reach about 10\% of the signal; as the ionization cloud spreads over at least a few \textmu{s} due to the combined effects of the photoelectron range and diffusion, afterpulsing can not be removed and its event-by-event fluctuations are not correctable.

\begin{figure}[H]
    \centering
    \includegraphics[width=12.5cm]{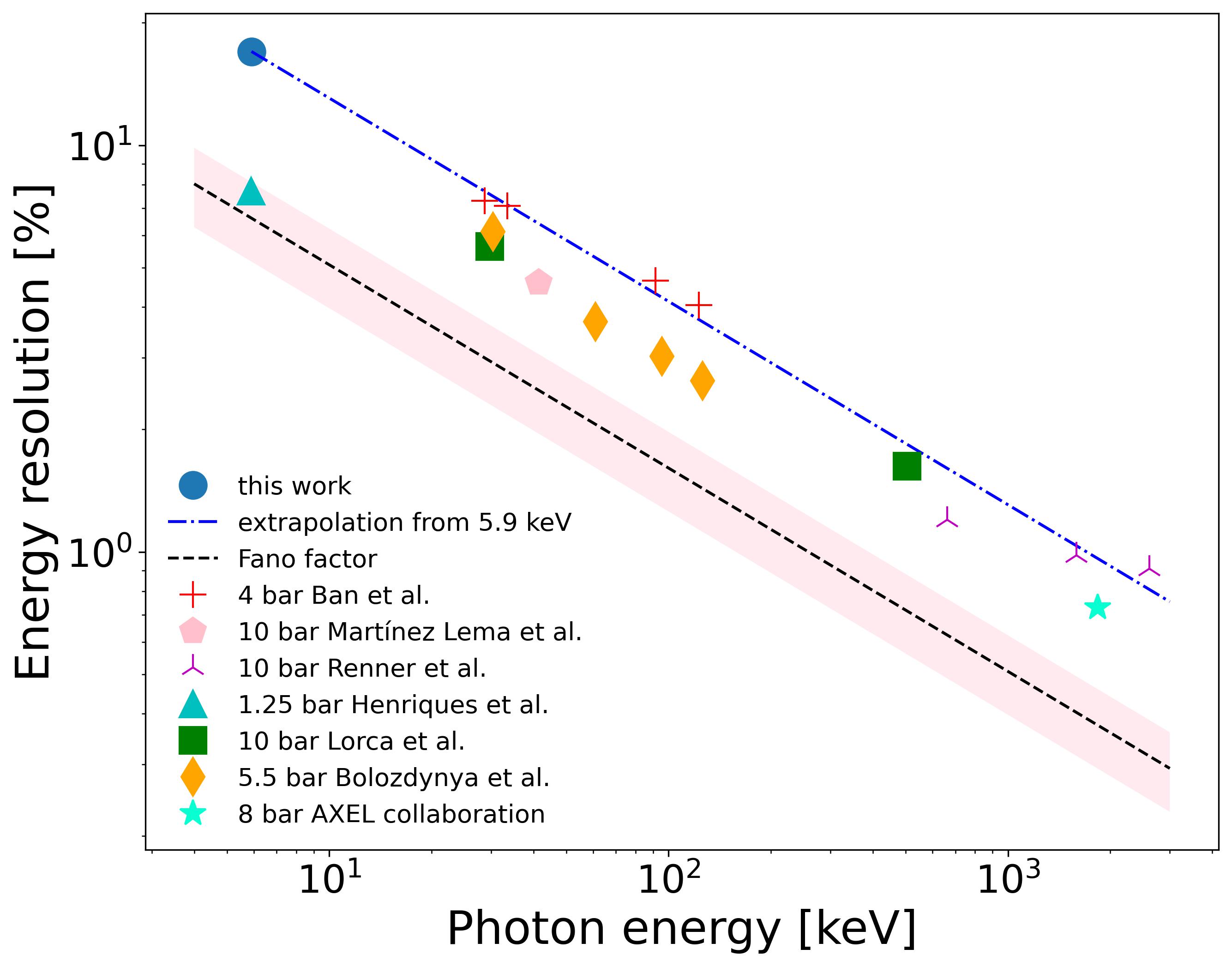}
     
    \centering
    \caption{\footnotesize Energy resolution in this work (dark-blue circle) and comparison with other experiments (Figure adapted from \cite{ChapterAngela}, with data from \cite{AXELSLAC, ER1, ER2, NEXTER, ER4, ER5, ER6}). The Fano factor represented by the dashed line is an average value from~\cite{FANO1, FANO2, FANO3, FANO4, FANO5, FANOLAST}, while the band shows the region comprising the highest and lowest values, considering the entire energy range involved. }
    \label{fig:comparison}
\end{figure}

Notwithstanding, an extrapolation of present results up to the $Q_{\beta\beta}$-scale of $^{136}$Xe (2.45~MeV) indicates that the achieved energy resolution is competitive with the ones reached by current experiments and other R\&D efforts (Figure~\ref{fig:comparison}). The recent result of 0.73\% at 1.836~MeV (star), reported for Teflon-based perforated structures in \cite{AXELSLAC}, gives further support to the notion that, in practical applications, this type of EL-structures are already within the double-mesh performance limit in terms of optical gain and energy resolution.

\begin{figure}[H]
    \centering
    \includegraphics[width=12.5cm]{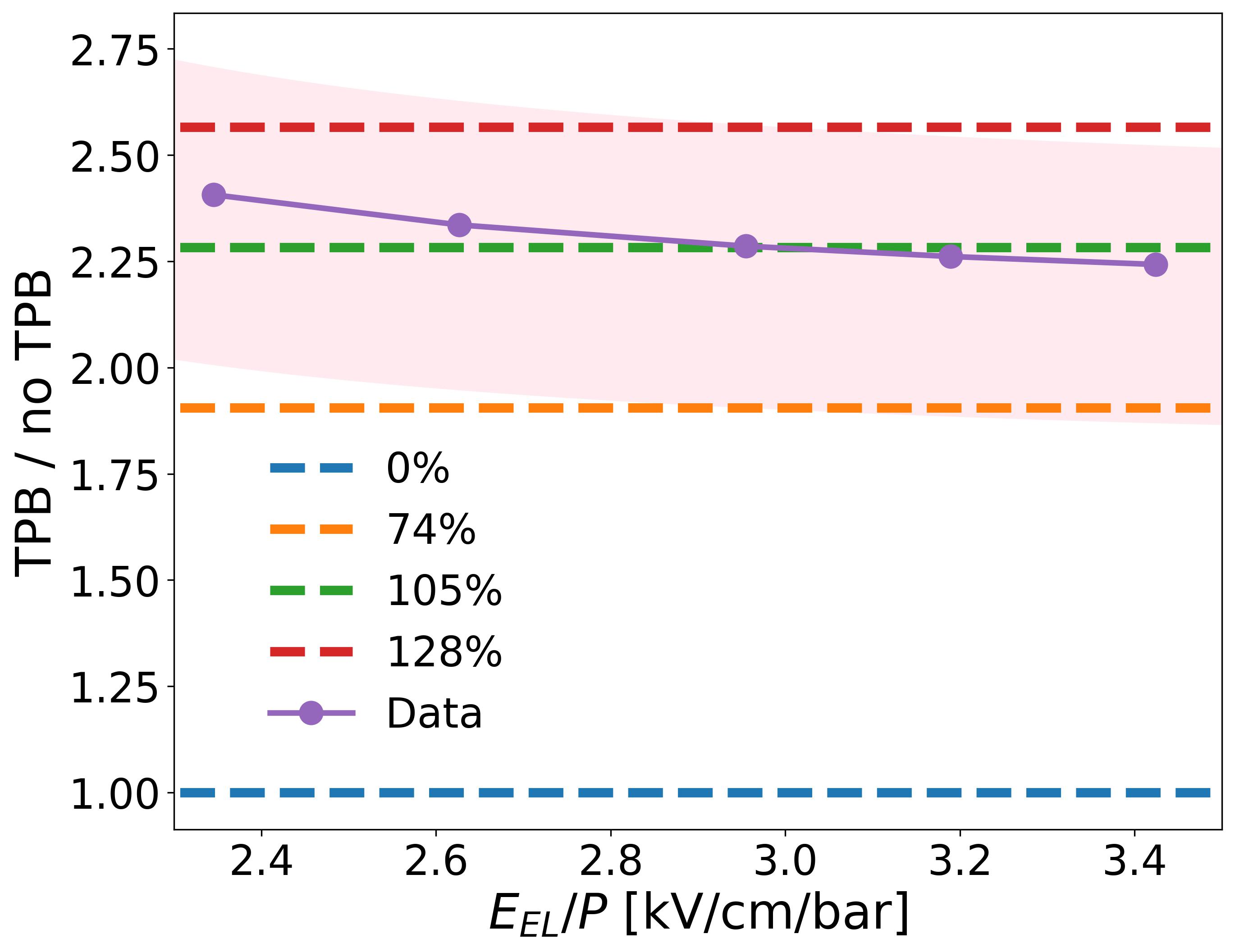}
     
    \centering
    \caption{\footnotesize Ratio of the light yield of the TPB-coated FAT-GEM to the uncoated one (purple circles), as a function of the pressure-reduced electric field, obtained in xenon at 4~bar. The light-red band accounts for the systematic uncertainty in data. Wavelength-shifting-efficiency (WLSE) has been estimated by comparison with the ratio predicted in simulations (WLSE values indicated by dashed lines). }
    \label{fig:wrapupsimu}
\end{figure}

To better understand the present technology limits, it is relevant to estimate the wavelength-shifting-efficiency (WLSE) achieved for the TPB-coating process inside the structure channels. The ratio of the TPB-coated structure to the non-coated one has been used for this purpose, as the ratio should be more immune to mis-modeling effects (Figure \ref{fig:wrapupsimu}).
The uncertainty band was taken to be 15\%. As shown, the estimated WLSE of the TPB lies within the 74\% to 128\% range, centered at 105\%, in line with the literature results~\cite{lars,tpbeff}. Simulations performed for the structure with ESR produced slightly lower values of the WLSE of around 94\%, that could point to a bad optical coupling (or modeling) of the ESR reflector.

\begin{figure}[H]
    \centering
    \includegraphics[width=12.5cm]{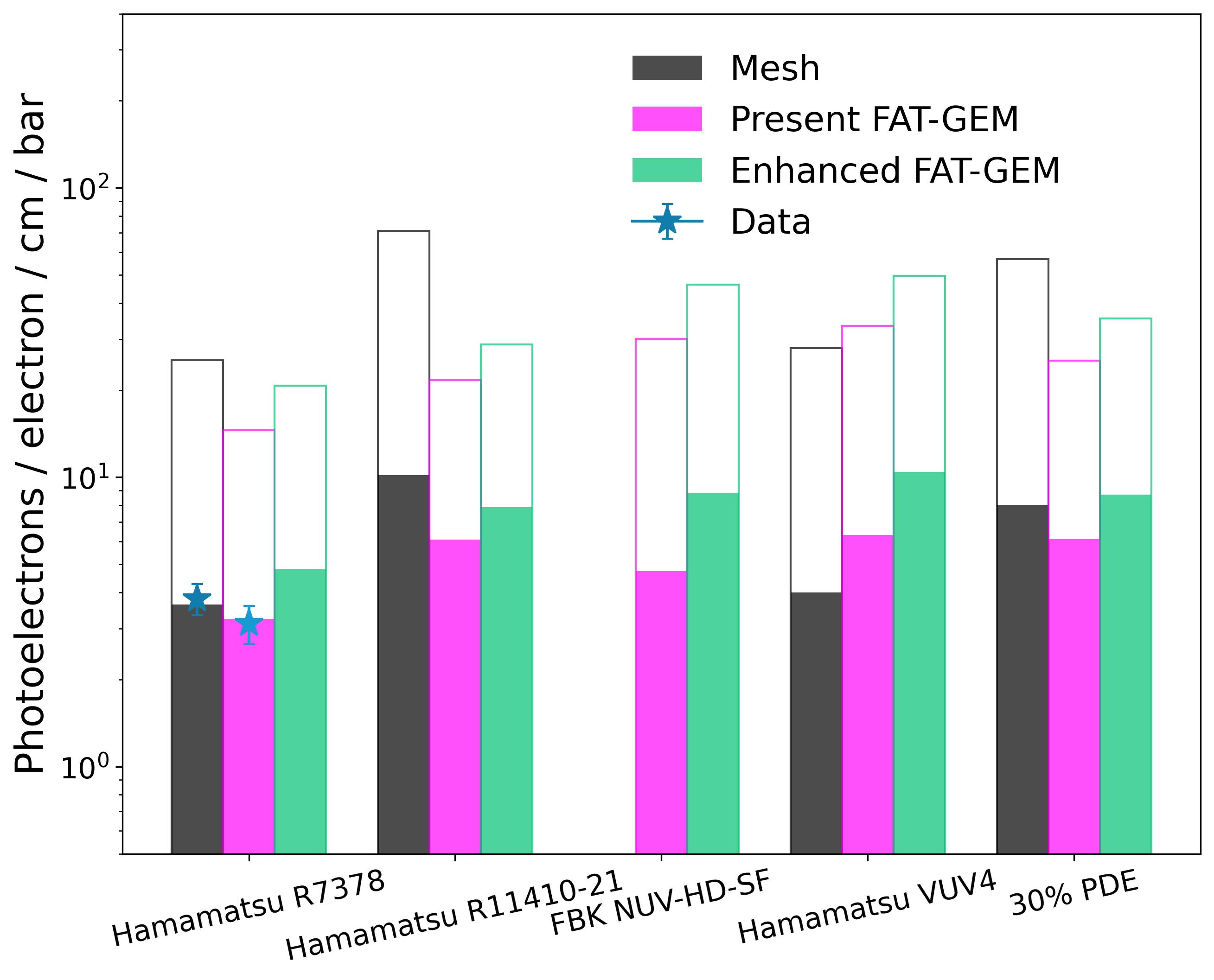}
    \centering
    \caption{\footnotesize Simulated light yields in phe/e/cm/bar when considering the PDE's of different photosensors, for a pressure-reduced field of 3.65~kV/cm/bar, $P=4$~bar and a gas gap of 5~mm. The empty bars refer to a half-hemisphere sensor, while the full bars assume the geometry in this work (2.1 cm diameter photosensor at 1.5~cm distance). The structures considered are a double-mesh (black), the FAT-GEM developed in this work (pink), and an `enhanced' FAT-GEM after the modifications discussed in text (green). The blue stars refer to experimental data in this work.}
    \label{fig:sensors}
\end{figure}

The process of wavelength-shifting results in a different spectral content for the FAT-GEM and double-mesh configurations, so it is interesting to 
perform a comparison for different photosensors, based on the results achieved in this work. Figure \ref{fig:sensors} shows the simulated number of photoelectrons (normalized to cm$\cdot$bar) assuming the nominal PDE of each of the four exemplary photosensors discussed here (meshes in black, FAT-GEM (present) in pink, FAT-GEM (enhanced) in green). Moreover, a flat-response light sensor with 30\% PDE is also shown. The `enhanced' version of the FAT-GEM assumes a diffuse reflector instead of ESR to minimize the effect of photon trapping due to internal reflection inside the PMMA. Experimental data obtained in this work are given by stars ($P=4~$bar). All comparisons have been made at a pressure-reduced electric field of $E_{EL} = 3.65$~kV/cm/bar for which discharge-free operation of the FAT-GEM was comfortably achieved. Extrapolations to a half-hemisphere photosensor plane are shown by the empty bar.

\section{Conclusions} \label{sec:con}

The use of `ad hoc' perforated structures for electroluminescence is reaching maturity. Here we present performances for the FAT-GEM technology (Field-Assisted Transparent Gaseous Electroluminescence Multipliers), that are already matching those achievable for double-meshes, our proxy for the `ideal' uniform-field situation. The optical gain achieved in xenon at 4~bar, at a pressure-reduced electric field of 3.65~kV/cm/bar, is within 18\% of the value achieved in a double-mesh configuration with the same photosensor coverage, a bit over the systematic uncertainty of present data (15\%). Further, a good description of the observed yields can be achieved if assuming a WLSE of 105\% for the TPB-coated channels. The structure is electrically stable at 3.65~kV/cm/bar (4~bar) and 2.8~kV/cm/bar (10~bar), the latter limited by the maximum power supply of our equipment. Despite being in a high-field region, no systematic TPB degradation was observed during 20~h of continuous operation.
Assuming complete hemispherical coverage, the above WLSE figure extrapolates to a number of detected photoelectrons (at 4~bar) in the range of: 29(41) for Hamamatsu R7378 PM, 43(57) for Hamamatsu R11410-21 PM, 60(92) for FBK NUV-HD-SF SiPM and 66(99) for Hamamatsu VUV4 SiPM when considering the present (or enhanced) FAT-GEM structures. Higher values are anticipated at higher pressures. The observed energy resolution, on the other hand, extrapolates to 0.84\% (0.91\%) FWHM for 4~bar (10~bar) at the $Q_{\beta\beta}$ of $^{136}$Xe, competitive for next-generation $\beta\beta0\nu$ experiments. Evidence has been presented, pointing to the intrinsic resolution of the structure being better than that.

The FAT-GEM concept combines the properties of gas scintillation and light-guides, being radiopure, versatile in design and intrinsically transparent from the hard VUV up to visible and near-IR regions (1600~nm). It is called to solve the scaling issues for large-volume chambers and perhaps offer an universal technique to implement electroluminescence readouts in noble-elements.

\section{Acknowledgments}

This research has been sponsored by RD51 funds through its `common project' initiative, and has received financial support from Xunta de Galicia (Centro singular de investigación de Galicia accreditation 2019-2022), and by the “María de Maeztu” Units of Excellence program MDM-2016-0692. DGD was supported by the Ram\'on y Cajal program (Spain) under contract number RYC-2015-18820.
 This research was also partly funded by the Spanish Ministry (‘Proyectos de Generación de Conocimiento’, PID2021-125028OB-C21), the National Science Centre, Poland (Grant No. 2019/03/X/ST2/01560), the International Research Agenda Programme AstroCeNT (MAB/2018/7) funded by the Foundation for Polish Science from the European Regional Development Fund~(ERDF), and the European Union’s Horizon~2020 research and innovation programme under grant agreement No~952480 (DarkWave project). The prototype construction was carried out with the use of CEZAMAT (Warsaw) cleanroom infrastructures financed by the ERDF; we thank Maciej Trzaskowski and CEZAMAT staff for technical support. We thank Carlos Guerra (Spartech/Polycast) for kindly providing the acrylic samples for this work; as usual, expert screening work by I. Bandac (Laboratorio Subterráneo de Canfranc) was essential to the radiopurity results presented here.

\bibliographystyle{elsarticle-num}

\end{document}